\documentclass[final,3p,times,twocolumn]{elsartm}
\biboptions{comma,sort&compress}
\usepackage{graphicx}

\newcommand{\RTau}[1]{R_{\tau, \mbox{\tiny #1}}}
\newcommand{\DeltaQCD}[1]{\Delta^{\mbox{\tiny #1}}_{\mbox{\tiny QCD}}}
\newcommand{\ind}[2]{^{\mbox{\scriptsize $#1$}}_{\mbox{\scriptsize #2}}}
\newcommand{\inds}[2]{^{\mbox{\tiny $#1$}}_{\mbox{\tiny #2}}}
\def\Vud{V_{\mbox{\scriptsize ud}}}
\def\Sew{S_{\!\mbox{\tiny EW}}}
\def\dpew{\delta'_{\mbox{\tiny EW}}}
\def\MTau{M_{\tau}}
\def\ML{M_{l}}
\def\Nc{N_{\mbox{\scriptsize c}}}
\def\nf{n_{\mbox{\scriptsize f}}}
\def\va{_{\mbox{\tiny V/A}}}
\def\zva{\zeta\va}

\begin{document}

\begin{frontmatter}

\title{Hadronic effects in low--energy QCD: inclusive $\tau$~lepton decay}
\author{A.V.~Nesterenko}
\address{Bogoliubov Laboratory of Theoretical Physics,
Joint Institute for Nuclear Research,
Joliot Curie 6, Dubna, Moscow region, 141980, Russia}
\ead{nesterav@theor.jinr.ru}

\begin{abstract}
\noindent
The inclusive $\tau$~lepton hadronic decay is studied within Dispersive
approach to QCD. The significance of effects due to hadronization is
convincingly demonstrated. The approach on hand proves to be capable of
describing experimental data on $\tau$~lepton hadronic decay in vector and
axial--vector channels. The vicinity of values of QCD scale parameter
obtained in both channels bears witness to the self--consistency of
developed approach.
\end{abstract}


\end{frontmatter}

The inclusive $\tau$~lepton hadronic decay provides a clean environment
for the study of the nonperturbative aspects of Quantum
Chromodynamics~(QCD) at low energies. In particular, this process is
commonly employed in tests of~QCD and entire Standard Model, that, in
turn, furnishes stringent constraints on possible New Physics beyond
the latter.

\medskip

The relevant experimentally measurable quantity is the ratio of the total
width of $\tau$~lepton decay into hadrons to the width of its leptonic
decay, which can be decomposed into several parts:
\begin{eqnarray}
\label{RTauGen}
R_{\tau} \!\!\!\!&=&\!\!\!\! \frac{\Gamma(\tau^{-} \to \mbox{hadrons}^{-}\, \nu_{\tau})}
{\Gamma(\tau^{-} \to e^{-}\, \bar\nu_{e}\, \nu_{\tau})}
\nonumber \\[1mm]
\!\!\!\!&=&\!\!\!\!
\RTau{V}^{\mbox{\tiny $J$=0}} + \RTau{V}^{\mbox{\tiny $J$=1}} +
\RTau{A}^{\mbox{\tiny $J$=0}} + \RTau{A}^{\mbox{\tiny $J$=1}} + \RTau{S}.
\end{eqnarray}
In the second line of this equation the first four terms account for the
hadronic decay modes involving light quarks~(u, d) only and associated
with vector~(V) and axial--vector~(A) quark currents, respectively,
whereas the last term accounts for the $\tau$~lepton decay modes which
involve strange quark. The superscript~$J$ indicates the angular momentum
in the hadronic rest frame.

\begin{figure*}[t]
\centerline{\includegraphics[width=75mm]{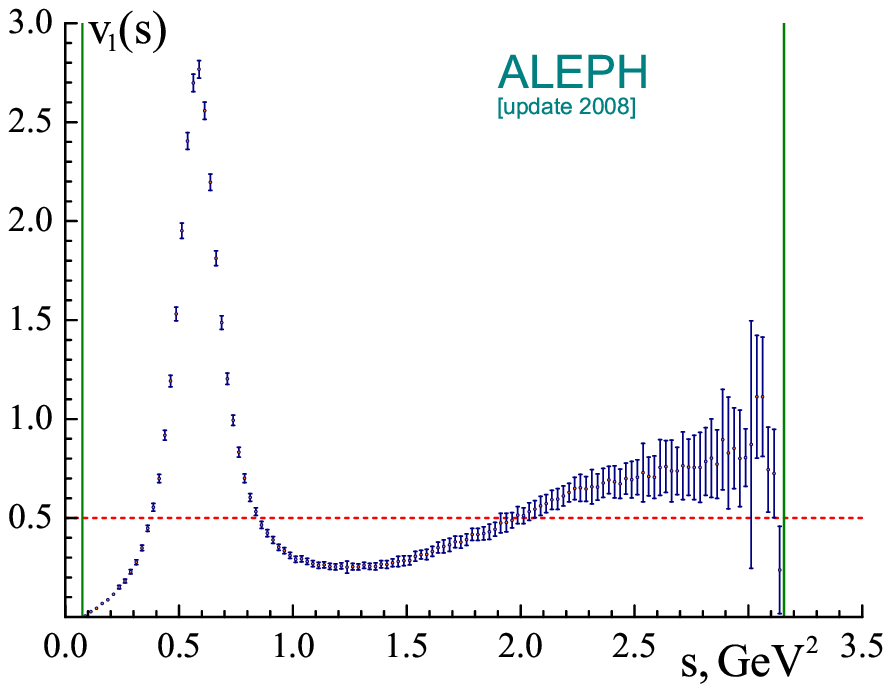}%
\hspace{11.7mm}%
\includegraphics[width=75mm]{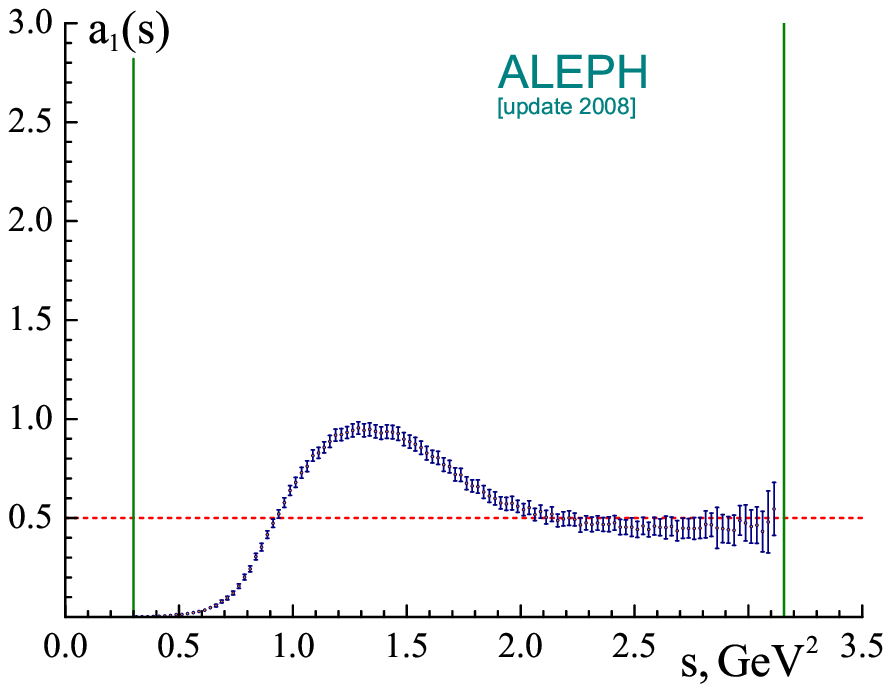}}
\caption{\scriptsize The inclusive vector and axial--vector spectral
functions~\cite{ALEPH08}. Vertical solid lines mark the boundaries of
respective kinematic intervals, whereas horizontal dashed lines denote
the naive massless parton model prediction.}
\label{Plot:SpFun}
\end{figure*}

All the quantities appearing in the second line of Eq.~(\ref{RTauGen}) can
be evaluated by making use of the spectral functions, which are determined
from the experiment. For the zero angular momentum (\mbox{$J=0$}) the
vector spectral function vanishes, whereas the \mbox{axial--vector} one is
usually approximated by Dirac $\delta$--function, since the main
contribution comes from the pion pole here. The experimental
predictions~\cite{ALEPH08} for the nonstrange spectral functions
corresponding to $J=1$ are presented in Fig.~\ref{Plot:SpFun}. In what
follows we shall restrict ourselves to the consideration of
terms~$\RTau{V}^{\mbox{\tiny $J$=1}}$ and~$\RTau{A}^{\mbox{\tiny $J$=1}}$
of $R_{\tau}$--ratio~(\ref{RTauGen}).

\medskip

The theoretical prediction for the aforementioned quantities reads
\begin{equation}
\RTau{V/A}^{\mbox{\tiny $J$=1}} = \frac{\Nc}{2}\,|\Vud|^2\,\Sew\,
\Bigl(\DeltaQCD{V/A} + \dpew \Bigr)\, ,
\end{equation}
where $\Nc=3$ is the number of colors, $|\Vud| = 0.97425 \pm 0.00022$ is
Cabibbo--Kobayashi--Maskawa matrix element~\cite{PDG2012}, $\Sew = 1.0194
\pm 0.0050$ and $\dpew = 0.0010$ stand for the electroweak corrections
(see Refs.~\cite{BNP, EWF}), and \looseness=-1
\begin{equation}
\label{DeltaQCDDef}
\DeltaQCD{V/A} = 2\int_{m\va^2}^{\ML^2}\!
f\biggl(\frac{s}{\ML^2}\biggr)\, R^{\mbox{\tiny V/A}}(s)\,
\frac{d s}{\ML^2}
\end{equation}
denotes the QCD contribution. In this equation
\begin{equation}
\label{RDef}
R^{\mbox{\tiny V/A}}(s) =
\frac{1}{\pi}\,\mbox{\normalfont Im}\!\lim_{\varepsilon \to 0_{+}}
\!\Pi^{\mbox{\tiny V/A}}(s + i \varepsilon),
\end{equation}
with~$\Pi^{\mbox{\tiny V/A}}(q^2)$ being the hadronic vacuum polarization
function, and $f(x) = (1-x)^{2}\,(1+2x)$.

In Eq.~(\ref{DeltaQCDDef}) $\ML$~denotes the mass of the lepton on hand,
whereas $m\va$~stands for the total mass of the lightest allowed hadronic
decay mode of this~lepton in the corresponding channel. The nonvanishing
value of~$m\va$ explicitly embodies the physical fact that $\tau$~lepton
is the only lepton which is heavy enough (\mbox{$\MTau \simeq
1.777\,$GeV~\cite{PDG2012}}) to decay into hadrons. Indeed, in the
massless limit ($m\va = 0$) the theoretical prediction for
$\DeltaQCD{}$~(\ref{DeltaQCDDef}) is nonvanishing\footnote{Specifically,
the leading--order term of Eq.~(\ref{DeltaQCDPert})
\mbox{$\Delta\inds{(0)}{pert}=1$} (which corresponds to the naive massless
parton model prediction for the Adler function~(\ref{AdlerPert})
$D\inds{(0)}{pert}(Q^2)=1$) does not depend on~$\ML$, and, therefore, is
unique for either lepton.} for either lepton (\mbox{$l = e, \mu, \tau$}),
whereas in the realistic case (\mbox{$m\va \neq 0$})
Eq.~(\ref{DeltaQCDDef}) acquires non--zero value for the case of the
$\tau$~lepton only. \looseness=-1

In general, it is convenient to perform the theoretical analysis
of inclusive $\tau$~lepton hadronic decay in terms of the Adler
function~\cite{Adler} (the indices ``V'' and~``A'' will only be
shown when relevant hereinafter)
\begin{equation}
\label{AdlerDef}
D(Q^2) = - \frac{d\, \Pi(-Q^2)}{d \ln Q^2}, \quad
Q^2 =-q^2=-s.
\end{equation}
Within perturbation theory the ultraviolet behavior of this function can
be approximated by power series in the strong running
coupling~$\alpha\ind{}{s}(Q^2)$: $D(Q^2) \simeq
D\ind{(\ell)}{pert}(Q^2)$ for $Q^2\to\infty$, where
\begin{equation}
\label{AdlerPert}
D\ind{(\ell)}{pert}(Q^2) = 1 + \sum\nolimits_{j=1}^{\ell}
d_{j} \Bigl[\alpha\ind{(\ell)}{pert}(Q^2)\Bigr]^{j}.
\end{equation}
At the one--loop level (i.e., for \mbox{$\ell=1$})
$\alpha\ind{(1)}{pert}(Q^2) = 4\pi/(\beta_{0}\,\ln z)$, $z=Q^2/\Lambda^2$,
$\beta_{0}=11-2\nf/3$, $\Lambda$~denotes the QCD scale parameter, $\nf$~is
the number of active flavors, and~$d_{1}=1/\pi$, see
papers~\cite{AdlerPert4Lab, AdlerPert4Lc} and references therein for the
details. In what follows the one--loop level with $\nf=3$ active flavors
will be assumed.

\medskip

For the beginning, let us study the massless limit, that implies that the
masses of all final state particles are neglected~($m=0$). By making use
of definitions~(\ref{RDef}) and~(\ref{AdlerDef}), integrating by parts,
and additionally employing Cauchy theorem, the
quantity~$\DeltaQCD{}$~(\ref{DeltaQCDDef}) can be represented as (see
Refs.~\cite{P91, DP92, BNP})
\begin{equation}
\label{DeltaQCDCauchy}
\DeltaQCD{} \!=\! \frac{1}{2\pi}\!\! \int_{-\pi}^{\pi}\!\!\!
D\bigl(M_{\tau}^{2}\,e^{i\theta}\bigr)
\bigl(1 \!+\! 2e^{i\theta} \!-\! 2e^{i3\theta} \!-\! e^{i4\theta}\bigr) d \theta.
\end{equation}

It is necessary to outline here that Eq.~(\ref{DeltaQCDCauchy}) can be
derived from Eq.~(\ref{DeltaQCDDef}) only for the massless limit of
``genuine physical'' Adler function~$D\ind{}{phys}(Q^2)$, which possesses
the correct analytic properties in the kinematic variable~$Q^2$. However,
in Eq.~(\ref{DeltaQCDCauchy}) one usually directly employs the
perturbative approximation~$D\ind{}{pert}(Q^2)$~(\ref{AdlerPert}), which
has unphysical singularities in~$Q^2$. At the \mbox{one--loop} level this
prescription eventually leads to
\begin{equation}
\label{DeltaQCDPert}
\Delta\ind{}{pert} \!=\! \Delta\ind{(0)}{pert} \!+\!
\frac{4}{\beta_{0}}\!\int_{0}^{\pi}
\frac{\lambda A_{1}(\theta)+\theta A_{2}(\theta)}{\pi(\lambda^2+\theta^2)}
\,d\theta,
\end{equation}
where
$A_{1}(\theta) = 1 + 2\cos(\theta) - 2\cos(3\theta) - \cos(4\theta)$,
$A_{2}(\theta) \!=\! 2\sin(\theta) - 2\sin(3\theta) - \sin(4\theta)$,
$\lambda \!=\! \ln \bigl( \MTau^2/\Lambda^2 \bigr)$,
and \mbox{$\Delta\ind{(0)}{pert} = 1$}.

\begin{figure*}[t]
\centerline{\includegraphics[width=75mm]{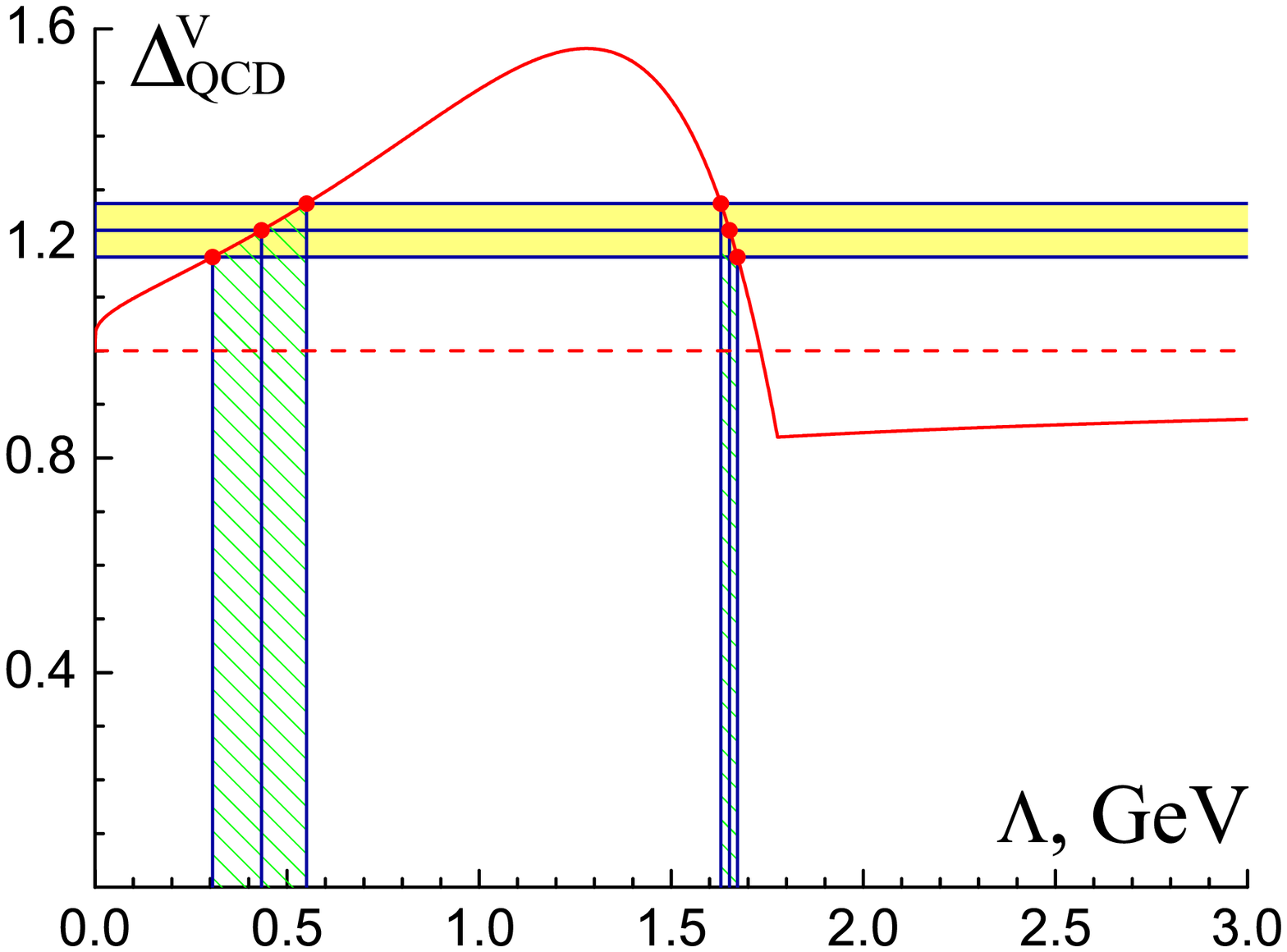}%
\hspace{11.7mm}%
\includegraphics[width=75mm]{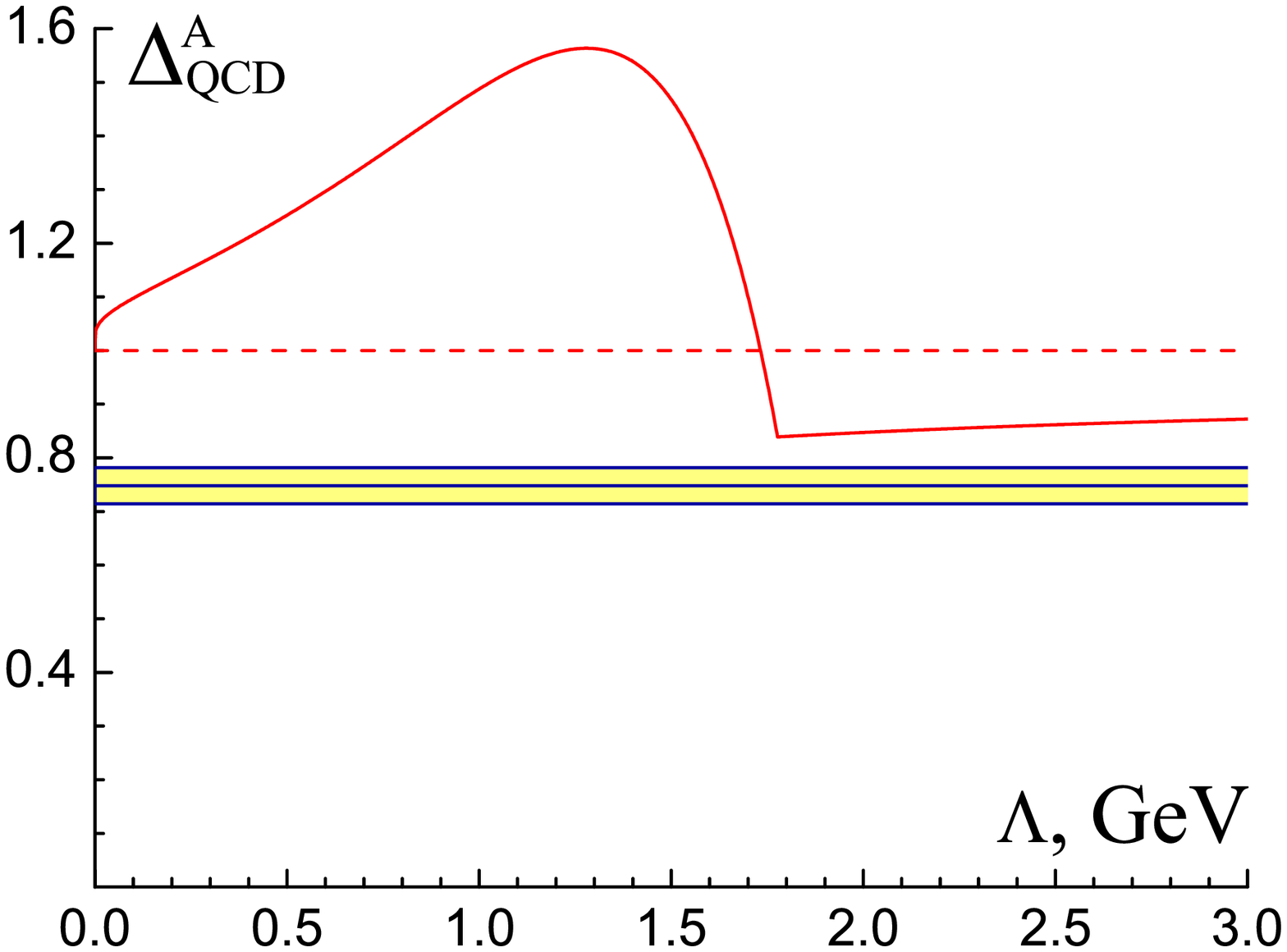}}
\caption{\scriptsize Comparison of the one--loop perturbative expression
$\Delta\inds{}{pert}$~(\ref{DeltaQCDPert}) (solid curves) with relevant
experimental data~(\ref{DeltaQCDExp}) (horizontal shaded bands). The
solution for QCD scale parameter~$\Lambda$ (if exists) is shown by
vertical dashed band.}
\label{Plot:Pert}
\end{figure*}

It is worth noting also that perturbative approach provides identical
predictions for functions~(\ref{DeltaQCDDef}) in vector and axial--vector
channels (i.e., \mbox{$\Delta\ind{\mbox{\tiny V}}{pert} \equiv
\Delta\ind{\mbox{\tiny A}}{pert}$}). However, their experimental
values~\cite{ALEPH9806, ALEPH08} are different, namely
\begin{equation}
\label{DeltaQCDExp}
\Delta\ind{\mbox{\tiny V}}{exp} \!\!=\! 1.224 \!\pm\! 0.050, \quad \!
\Delta\ind{\mbox{\tiny A}}{exp} \!\!=\! 0.748 \!\pm\! 0.034.
\end{equation}
The juxtaposition of these quantities with perturbative
result~(\ref{DeltaQCDPert}) is presented in Fig.~\ref{Plot:Pert}. As one
can infer from this figure, for vector channel there are two solutions for
the QCD scale parameter, namely,
$\Lambda=\bigl(434^{+117}_{-127}\bigr)\,$MeV and
$\Lambda=\bigl(1652^{+21}_{-23}\bigr)\,$MeV. Commonly, the first of these
solutions is retained, whereas the second one is merely disregarded.
As~for the axial--vector channel, the perturbative approach fails to
describe the experimental data on inclusive $\tau$~lepton hadronic decay,
since for any value of~$\Lambda$ the function
$\Delta\ind{}{pert}$~(\ref{DeltaQCDPert}) exceeds $\Delta\ind{\mbox{\tiny
A}}{exp}$~(\ref{DeltaQCDExp}). \looseness=-1

\medskip

It is crucial to emphasize that the presented above massless limit
completely leaves out the effects due to hadronization, which play an
important role in the studies of the strong interaction processes at low
energies. Specifically, the mathematical realization of the physical fact,
that in a strong interaction process no hadrons can be produced at
energies below the total mass~$m$ of the lightest allowed hadronic final
state, consists in the fact that the beginning of cut of corresponding
hadronic vacuum polarization function~$\Pi(q^2)$ in complex
\mbox{$q^2$--plane} is located at the threshold of hadronic
production~$q^2=m^2$, but not at~$q^2=0$. Such restrictions are inherently
embodied within relevant dispersion relation. In turn, the latter imposes
stringent physical nonperturbative constraints on the quantities on hand,
which should certainly be accounted for when one is trying to go beyond
the limitations of perturbation theory. \looseness=-1

Thus, the nonperturbative constraints, which dispersion
relation~\cite{Adler}
\begin{equation}
\label{AdlerDisp}
D(Q^2) = Q^2\! \int_{m^2}^{\infty}\!
\frac{R(s)}{(s + Q^2)^2} \, d s
\end{equation}
imposes on the Adler function~(\ref{AdlerDef}), have been merged with
corresponding perturbative result~(\ref{AdlerPert}) in the framework of
Dispersive approach to QCD, that has eventually led to the following
integral representations for functions~(\ref{RDef}) and~(\ref{AdlerDef})
(see Refs.~\cite{MAPT2, NPQCD07} for the details):
\begin{eqnarray}
\label{RMAPT}
R(s) \!\!\!\!&=&\!\!\!\! r\ind{(0)}{}(s) + \theta\Bigl(1-\frac{m^2}{s}\Bigr)
\!\int_{s}^{\infty}\!
\rho(\sigma)\, \frac{d \sigma}{\sigma}\, ,
\\
\label{AdlerMAPT}
D(Q^2) \!\!\!\!&=&\!\!\!\! d\ind{(0)}{}(Q^2) + \!\!
\int_{m^2}^{\infty}\!\!\! P(Q^2,m^2,\sigma)\, \rho(\sigma)\,
\frac{d \sigma}{\sigma}.
\end{eqnarray}
In these equations $\rho(\sigma)$ denotes the spectral density,
$P(Q^2,m^2,\sigma)=Q^2(\sigma - m^2)/((Q^2+m^2)(\sigma+Q^2))$, and
$\theta(x)$ is the unit step--function ($\theta(x)=1$ if $x \ge 0$ and
$\theta(x)=0$ otherwise). It is worth noting also that in the massless
limit ($m=0$) expressions~(\ref{RMAPT}) and~(\ref{AdlerMAPT}) become
identical to those of the \mbox{so--called} Analytic Perturbation
Theory~\cite{APTSS, APTMS} (see also Refs.~\cite{Cvetic, APT1, APT2}).
But, as it was mentioned above, it is essential to keep the hadronic
mass~$m$ nonvanishing within the approach on hand. \looseness=-1

\begin{figure*}[t]
\centerline{\includegraphics[width=75mm]{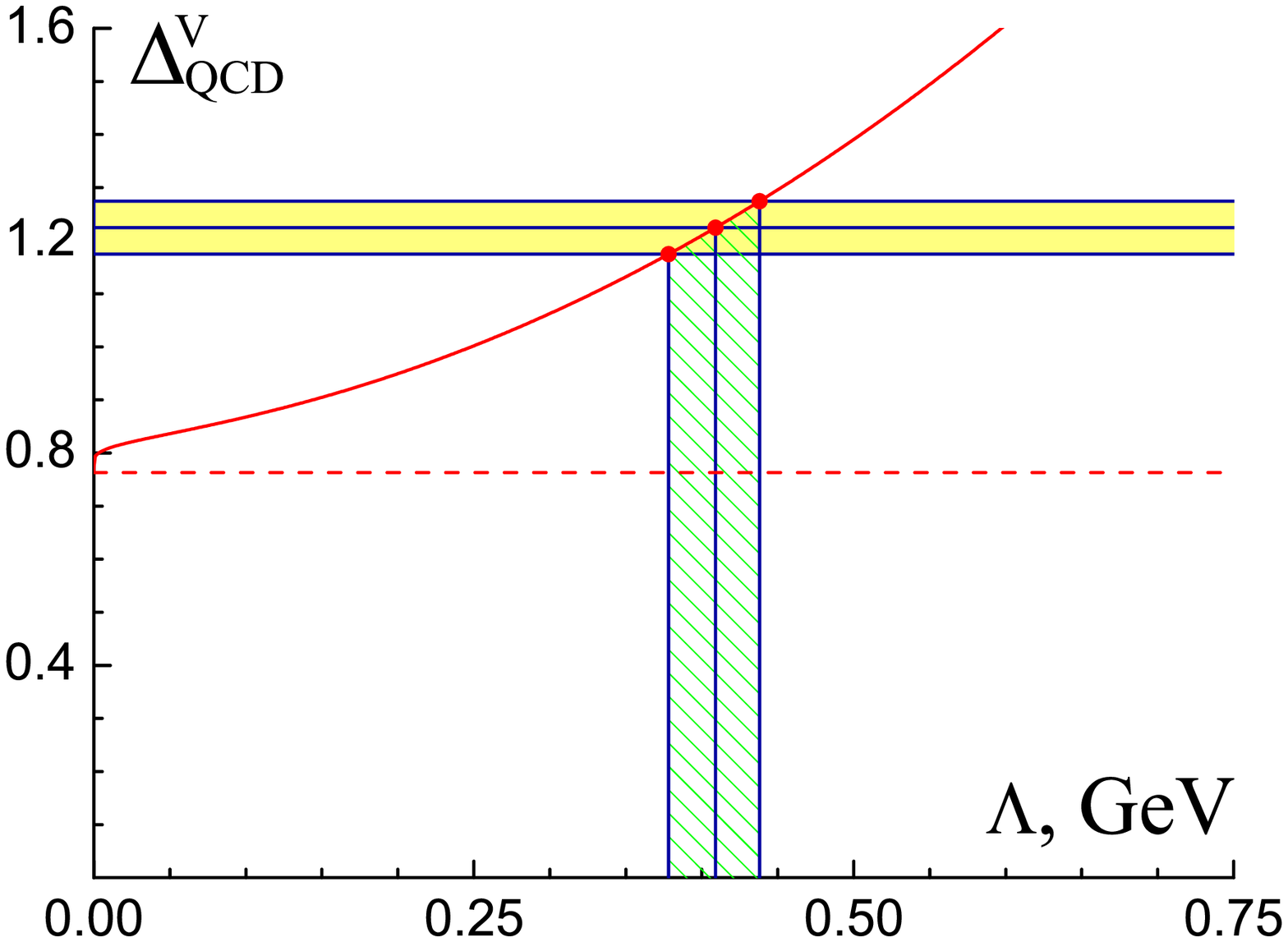}%
\hspace{11.7mm}%
\includegraphics[width=75mm]{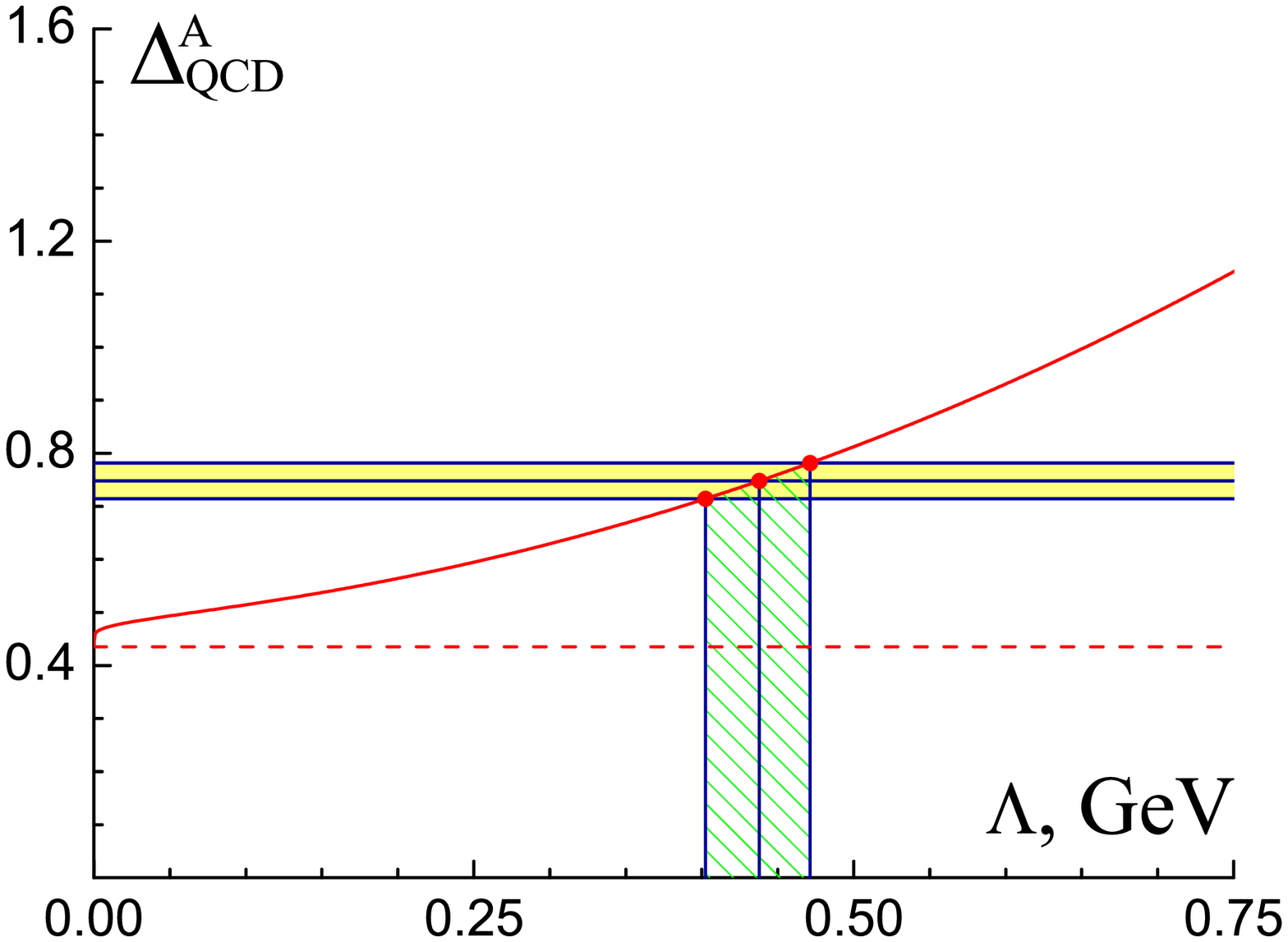}}
\caption{\scriptsize Comparison of expression
$\Delta\inds{\!\mbox{V/A}}{QCD}$~(\ref{DeltaQCD_MAPT_ST}) (solid curves)
with relevant experimental data~(\ref{DeltaQCDExp}) (horizontal shaded
bands). The solutions for QCD scale parameter~$\Lambda$ are shown by
vertical dashed bands.}
\label{Plot:MAPT_ST}
\end{figure*}

\medskip

Let us proceed now to the description of inclusive $\tau$~lepton hadronic
decay within Dispersive approach~\cite{MAPT2, NPQCD07} (see also
papers~\cite{PRD64, PRD71, PRD77} and references therein). In~this
analysis the effects due to hadronization will be retained (in other
words, the expressions~(\ref{RMAPT}) and~(\ref{AdlerMAPT}) will be used
instead of their perturbative approximations and the hadronic mass~$m$
will be kept nonvanishing). The so--called ``smooth kinematic threshold''
for the leading--order term of function~$R(s)$ (see, e.g.,
Refs.~\cite{Feynman, QCDAB}) will also be employed:
\begin{eqnarray}
r\ind{(0)}{\mbox{\tiny V/A}}(s) \!\!\!\!\!&=&\!\!\!\!\!
(1-m\va^{2}/s)^{3/2},
\\
d\ind{(0)}{\mbox{\tiny V/A}}(Q^2) \!\!\!\!\!&=&\!\!\!\!\! 1 + \frac{3}{\xi}
\biggl[\!1 + \frac{u(\xi)}{2}
\ln\Bigl[\!1 \!+\! 2\xi\bigl(1 \!-\! u(\xi)\bigr)\Bigr]\biggr] ,
\end{eqnarray}
where $u(\xi)=\sqrt{1+\xi^{-1}}$ and~$\xi=Q^2/m\va^{2}$. Besides, the
following model for the one--loop spectral density will be adopted:
\begin{equation}
\label{RhoDef}
\rho(\sigma) = \frac{4}{\beta_{0}}\frac{1}{\ln^{2}(\sigma/\Lambda^2)+\pi^2} +
\frac{\Lambda^2}{\sigma}\, ,
\end{equation}
see papers~\cite{Tau11, PRD62, Review} and references therein. The first
term in the right--hand side of Eq.~(\ref{RhoDef}) is the \mbox{one--loop}
perturbative contribution, whereas the second term represents
intrinsically nonperturbative part of the spectral density.

Eventually, all this has led to the following expression for the quantity
$\DeltaQCD{V/A}$~(\ref{DeltaQCDDef}) within Dispersive approach (see
Refs.~\cite{Tau11, Prep} for the details):
\begin{eqnarray}
\label{DeltaQCD_MAPT_ST}
\DeltaQCD{V/A} \!\!\!\!\!&=&\!\!\!\!\! \sqrt{1-\zva}\,
\biggl(1 + 6\zva - \frac{5}{8}\zva^{2}
+\frac{3}{16}\zva^{3}\biggr)
\nonumber \\
&&-3\zva \biggl(1 + \frac{1}{8}\zva^{2} - \frac{1}{32}\zva^{3} \biggr)
\nonumber \\
&&~\,\times\ln\biggl[\frac{2}{\zva}\Bigl(1+\sqrt{1-\zva}\Bigr)-1\biggr]
\nonumber \\
&&+ \int_{m\va^{2}}^{\infty}\!H\biggl(\frac{\sigma}{M_{\tau}^{2}}\biggr)\,
\rho(\sigma)\,\frac{d \sigma}{\sigma}\,,
\end{eqnarray}
where $H(x) = g(x)\,\theta(1-x) + g(1)\,\theta(x-1) - g(\zeta\va)$, $g(x)
= x (2 - 2x^2 + x^3)$, $m_{\mbox{\tiny V}}^{2} \simeq
0.075\,\mbox{GeV}^2$, $m_{\mbox{\tiny A}}^{2} \simeq 0.300\,\mbox{GeV}^2$,
and $\zeta\va = m\va^{2}/\MTau^{2}$.

The comparison of obtained result~(\ref{DeltaQCD_MAPT_ST}) with
experimental data~(\ref{DeltaQCDExp}) yields nearly identical solutions
for the QCD scale parameter~$\Lambda$ in both channels, see
Fig.~\ref{Plot:MAPT_ST}. Namely, $\Lambda = (408 \pm 30)\,$MeV for vector
channel and $\Lambda = (437 \pm 34)\,$MeV for axial--vector one.
Additionally, both these solutions agree with aforementioned perturbative
solution for vector channel.

\medskip

The author is grateful to A.~Bakulev, D.~Boito, M.~Davier, and S.~Menke
for the stimulating discussions and useful comments. Partial financial
support of grant JINR--12--301--01 is acknowledged.


\vspace*{-3.5mm}
\begin{thebibliography}{99}
\vspace*{-1.5mm}

\bibitem{ALEPH08} M.~Davier, S.~Descotes--Genon, A.~Hocker,
  B.~Malaescu, and Z.~Zhang,
  Eur.\ Phys.\ J.\ C~\textbf{56} (2008) 305.


\bibitem{PDG2012} J.~Beringer {\it et al.}
  [Particle Data Group Collaboration],
  Phys.\ Rev.\ D~\textbf{86} (2012) 010001.


\bibitem{BNP} E.~Braaten, S.~Narison, and A.~Pich,
  Nucl.\ Phys.\ B \textbf{373} (1992) 581.


\bibitem{EWF} W.J.~Marciano and A.~Sirlin,
  Phys.\ Rev.\ Lett.\ \textbf{61} (1988) 1815;
  E.~Braaten and C.S.~Li,
  Phys.\ Rev.\ D \textbf{42} (1990) 3888.


\bibitem{Adler} S.L.~Adler, Phys.\ Rev.\ D \textbf{10} (1974) 3714.


\bibitem{AdlerPert4Lab} P.A.~Baikov, K.G.~Chetyrkin, and J.H.~Kuhn,
  Phys.\ Rev.\ Lett.\ \textbf{101} (2008) 012002;
  \textbf{104} (2010) 132004.


\bibitem{AdlerPert4Lc} P.A.~Baikov, K.G.~Chetyrkin, J.H.~Kuhn,
  and J.~Rittinger,
  Phys.\ Lett.\ B \textbf{714} (2012) 62.


\bibitem{P91} A.A.~Pivovarov,
  Z.\ Phys.\ C \textbf{53} (1992) 461.


\bibitem{DP92} F.~Le Diberder and A.~Pich,
  Phys.\ Lett.\ B \textbf{286} (1992) 147;
  \textbf{289} (1992) 165.


\bibitem{ALEPH9806} R.~Barate \textit{et al.} [ALEPH Collaboration],
  Eur.\ Phys.\ J.\ C~\textbf{4} (1998) 409;
  M.~Davier, A.~Hocker, and Z.~Zhang,
  Rev.\ Mod.\ Phys.\ \textbf{78} (2006) 1043.


\bibitem{MAPT2} A.V.~Nesterenko and J.~Papavassiliou,
  J.\ Phys.\ G \textbf{32} (2006) 1025.


\bibitem{NPQCD07} A.V.~Nesterenko, SLAC eConf C0706044 (2008) 25;
  arXiv:0710.5878~[hep-ph];
  Nucl.\ Phys.\ B (Proc.\ Suppl.) \textbf{186} (2009) 207.


\bibitem{APTSS} D.V.~Shirkov and I.L.~Solovtsov, Phys.\ Rev.\ Lett.\
  \textbf{79} (1997) 1209; Theor.\ Math.\ Phys.\ \textbf{150} (2007) 132.


\bibitem{APTMS} K.A.~Milton and I.L.~Solovtsov, Phys.\ Rev.\ D \textbf{55}
  (1997) 5295; \textbf{59} (1999) 107701.


\bibitem{Cvetic} G.~Cvetic and C.~Valenzuela,
  Braz.\ J.\ Phys.\ \textbf{38} (2008) 371.


\bibitem{APT1}
  K.A.~Milton, I.L.~Solovtsov, and O.P.~Solovtsova,
  Mod.\ Phys.\ Lett.\ A \textbf{21} (2006) 1355;
  G.~Cvetic, C.~Valenzuela, and I.~Schmidt,
  Nucl.\ Phys.\ B (Proc.\ Suppl.) \textbf{164} (2007) 308.


\bibitem{APT2}
  A.P.~Bakulev,
  Phys.\ Part.\ Nucl.\ \textbf{40} (2009) 715;
  G.~Cvetic and A.V.~Kotikov,
  J.\ Phys.\ G \textbf{39} (2012) 065005.


\bibitem{PRD64} A.V.~Nesterenko, Phys.\ Rev.\ D \textbf{64} (2001) 116009.


\bibitem{PRD71} A.V.~Nesterenko and J.~Papavassiliou,
  Phys.\ Rev.\ D \textbf{71} (2005) 016009.


\bibitem{PRD77} M.~Baldicchi, A.V.~Nesterenko, G.M.~Prosperi, and C.~Simolo,
  Phys.\ Rev.\ D \textbf{77} (2008) 034013.


\bibitem{Feynman}
  R.P.~Feynman, \textit{Photon--hadron interactions},
  Reading,~MA: Benjamin (1972) 282p.


\bibitem{QCDAB}
  A.I.~Akhiezer and V.B.~Berestetsky,
  \textit{Quantum electrodynamics},
  Interscience,~NY (1965) 868p.


\bibitem{Tau11} A.V.~Nesterenko, SLAC eConf C1106064 (2011) 23;
  arXiv:1106.4006~[hep-ph]; arXiv:1110.3415~[hep-ph].


\bibitem{PRD62} A.V.~Nesterenko, Phys.\ Rev.\ D \textbf{62} (2000) 094028.


\bibitem{Review} A.V.~Nesterenko,
  Int.\ J.\ Mod.\ Phys.\ A \textbf{18} (2003) 5475;
  Nucl.\ Phys.\ B (Proc.\ Suppl.) \textbf{133} (2004) 59.


\bibitem{Prep} A.V.~Nesterenko, in preparation.

\end{thebibliography}
\end{document}